# Estimation of Tissue Deformation and Interactive Forces in Robotic Surgery through Vision-based Learning

Srikar Annamraju, Yuxi Chen, Jooyoung Lim, Paul M Jeziorczak, Inki Kim, *Member, IEEE*

*Abstract*— *Goal:* A limitation in robotic surgery is the lack of force feedback, due to challenges in suitable sensing techniques. To enhance the surgeon's perception and precise force rendering, estimation of these forces along with tissue deformation level is presented here. *Methods:* An experimental test bed is built for studying the interaction, and the forces are estimated from the raw data. Since tissue deformation and stiffness are non-linearly related, they are independently computed for enhanced reliability. A Convolutional Neural Network (CNN) based vision model is deployed, and both classification and regression models are developed. *Results:* The forces applied on the tissue are estimated, and the tissue is classified based on its deformation. The exact deformation of the tissue is also computed. *Conclusions:* The surgeons can render precise forces and detect tumors using the proposed method. The rarely discussed efficacy of computing the deformation level is also demonstrated.

*Impact Statement*— While robotic surgery currently offers only visual feedback, the proposed method estimates the force applied by the robot on the patient, and also computes the deformation level on their tissue.

## I. INTRODUCTION

ROBOTIC surgeries are clinically proven to enhance the surgeon's experience and comfort, minimize incision on patient's body and thus reduce the healing period, as compared to the traditional surgery[1]. With ~1.5 million surgeries/year, their utility is increasing by ~15%/year[2]. This growth is attributed to its stunning accuracy and real-time visual feedback the surgeon perceives. However, a few of its limitations concern researchers to address these challenges. Apart from the implementation/logistic concerns over high cost, longer learning curves etc.[3], the technical limitations of the surgical robot are discussed in [4]. Though the field of haptics is thoroughly studied for decades, surgical robots does not yet incorporate force feedback, limiting surgeons in perceiving the patient's tissues (or other organs).

Physical perception of the tissue is an experience through which surgeon often infers the degree of healthiness of the affected organ. The changes in tissue stiffness and deformation are usually linked with musculoskeletal disorders and other pain conditions [5]. For instance, a tissue which feels harder than a usual one could be affected by tumor. However, computation of the tissue deformation levels have been seldom discussed in the state-of-art, as compared to the force estimation.

A significant reason (other reasons in 'SupplementaryMaterial') for not incorporating force feedback is the challenge of an appropriate sensing method. In the recent years, tactile sensors such as [7] are being developed to suit this purpose. However, even with an acceptable sensor, the choice of sensor placement brings further complications [8]. For reliable accuracy, the force sensor needs to be placed at the tool-tip, thus needing the sensor to be biodegradable and small enough, since it must then surpass the point of incision. Placing the sensor outside the patient's body circumvents this challenge but the external noises make the data unreliable. Thus, force *estimation* techniques are being explored by researchers as alternatives.

### A. Research Objectives

Without necessarily using an explicit force sensor, external force estimation is feasible [9], and *theoretically*, tissue deformation can be inferred from the same. But such an indirect method of inferring deformation from interaction force can become unreliable considering the non-linear dynamic properties of the tissue. For every distinct tissue, the range of elastic moduli remarkably varies over a significant range. Particularly given that significant robotic surgeries are laparoscopic, the abdominal tissues are one of the softest in the human body, next only to the neural tissues [10].

Thus, an attempt is made in this work to independently a) estimate the interactive force between the robotic tool-tip and the tissue, and b) estimate the deformation level of the tissue.

This work is supported by the Jump ARCHES program under project P384.

Dr. Srikar Annamraju is with the Coordinated Science Laboratory, University of Illinois Urbana Champaign. Yuxi Chen is with the ZJU-UIUC Institute, Zhejiang University. Jooyoung Lim is with the Department of Industrial and Enterprise Systems Engineering, University of Illinois Urbana Champaign. Dr. Paul Jeziorczak is with the Department of Clinical Surgery, OSF Saint Francis Medical Center, Peoria, IL, USA. Dr. Inki Kim is with the Department of Industrial and Enterprise Systems Engineering, University of Illinois Urbana Champaign. (Correspondence Email inkikim@illinois.edu).

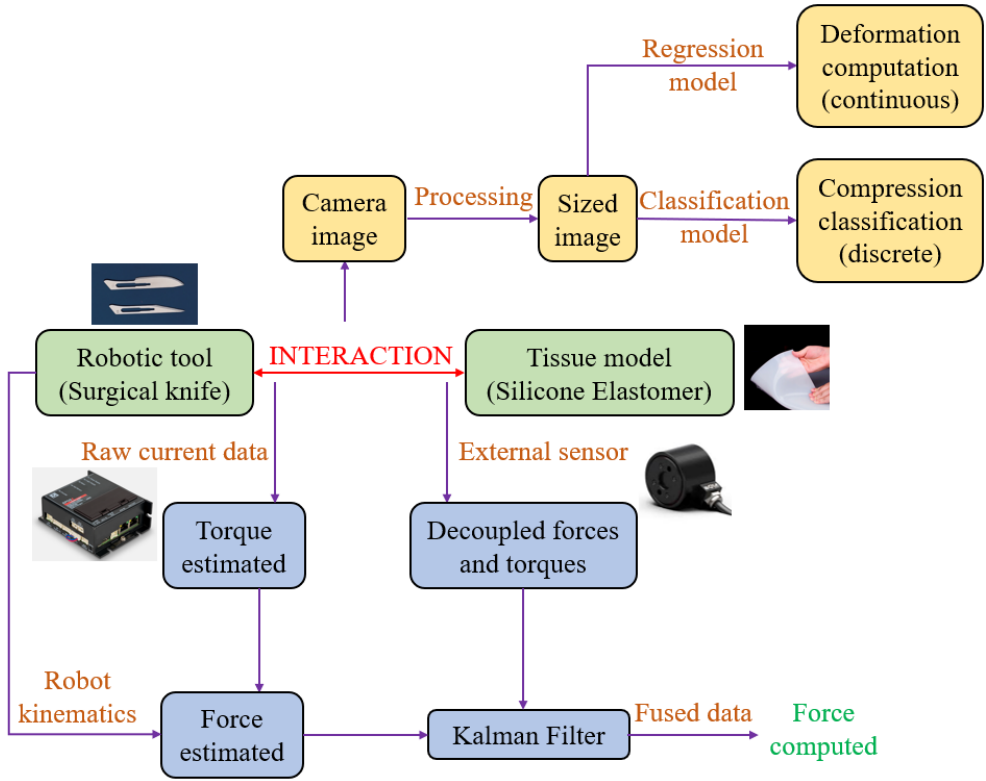

Figure 1. Layout of the Force and Deformation estimation scheme

Each of these objectives are carried out in a two-fold manner for enhanced accuracy.

*Force Estimation:* The ability to precisely render the desired force on the patient body is as important as making minimal incisions, and the current work helps the surgeons in rendering such precise forces. A prototype of the robotic tool-tissue interaction is developed, and the forces are first estimated through the motor currents. A high precision force sensor is then placed at an optimal location on the tool, and its data is fused with the prior estimates (from motor currents) through Kalman filtering. The fusion of data from two independent sources reduces the covariance from each measurement and increases the accuracy.

*Deformation Computation:* For computing the deformation level of the tissue, a camera is integrated to capture the incision of the tool-tip into the tissue model of the prototype. The observed compression levels are utilized to train an artificial intelligence (AI) algorithm which can classify the tissue into specific classes of compression. To further enable continuous (non-discrete) estimation, a regressor model is developed using CNNs. Both the modules of real-time object identification training and the regressor model adds its own advantage in identifying the deformation level of the tissue.

The objectives of this work are listed in 'SupplementaryMaterial' and is pictorially represented in Figure 1.

## II. MATERIALS AND METHODS

### A. Development of Prototype

A 2-DoF robotic tool-tissue interaction prototype is developed, with perpendicular motions to the tissue model and the tool. Both linear motions are achieved through rotary motors translated by a rack and pinion mechanism (Figure 2(a)).

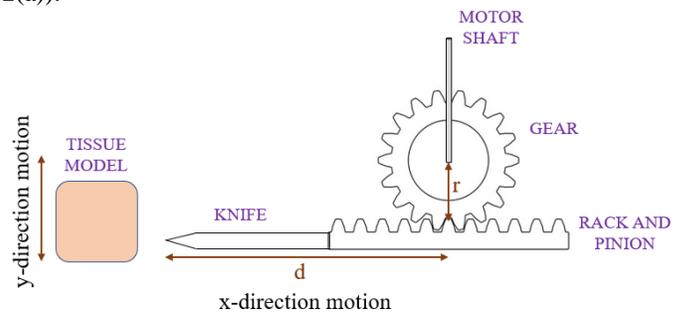

(a) Kinematic sketch of the interaction

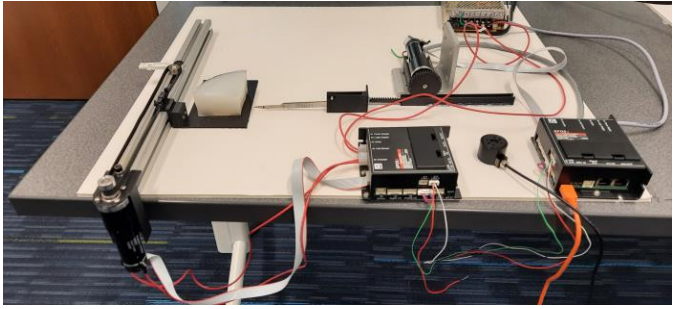
(b) Hardware of the prototype

Figure 2. Experimental Setup

The tool is a surgical knife, and the tissue model is a silicone elastomer, EcoFlex® (Smooth-On). Various tissue models are made with different consistencies of EcoFlex 10,30,40. A recent study [11] shows how effectively elastomers resemble tissue models, and their mathematical modelling.

The DC motors (RE25, Maxon®) are particularly chosen to be of low inertia (9.49gcm$^2$) so that the torque measurements do not interfere with the force estimation. The motor shaft is attached to a gear mechanism of 1:10 ($r=10$ in Figure2(a)) to multiply the torque potential of the end-effector (the surgical knife), so that even hard substances (such as tumors) in the tissue can be accurately realized. High precision 1024cpt encoders, with resolution of (1/4096$^{th}$)rotation are utilized, thus detecting every 1.75µm.

The need for precise control is met with Maxon EPOS-4(50V/5A) controllers, which can read the instantaneous motor currents in real time. These currents are translated into motor torques using the motor's torque constant. From the torques, the interaction forces at the tool-tip are computed using the instantaneous radial distance (variable *'d'* in Figure2(a)).

The EPOS are connected in a daisy chain through CAN communication, making the channel delays negligible in the order of nano-seconds. The motors and controllers are interfaced in real-time through LabView® (front panel of the interface shown in 'SupplementaryMaterial'). The experimental setup depicting the power supply, controllers, motors, tool and the tissue model is shown in Figure 2(b).

Whenever the knife probes into an environment, the current drawn by the motor spikes proportional to the environmental stiffness. Thus, the raw current data enables estimation of interaction forces. To further enhance the reliability of the estimate, an external sensor is incorporated.

### B. External Force Sensing and Fusion

The sensor used for augmenting the raw data is Mini-ONE from Bota Systems®, which is a 6-axis force-torque sensor. The sensor's force range of (0-50)N and a torque range of (0-1)Nm is high enough for detecting the penetrations in the tissue model, even accounting for the hard tumor-like interferences. It's unwanted crosstalk to <0.1% enables precise decoupling of forces across axes. The radius of the Mini-ONE being only 15mm enables its integration onto the tool, without obstructing the interaction and still not far from the point of incision (shown in 'SupplementaryMaterial').

### C. Camera Integration

The approach used in this work to detect the deformation level for the tissue is to develop a vision model and perform AI-based classification. A recent review [12] shows the shift of estimation techniques from tissue properties to learning-based approaches reducing the computational complexity. It is also demonstrated that the combination of vision model with the tool data yields the best force estimation quality [13]. In this view, the approach presented here is a novel means of integrating image-based learning with measured tool position. Thus, a hybrid of data driven and learning approaches is adopted.

A 5MP 6MM C-mount lens from Get-cameras® is installed 30cm above the interaction (shown in 'SupplementaryMaterial'). The camera's high resolution and a distortion < 0.1% makes it an ideal choice for capturing fine changes in the deformation.

The deformation prediction problem is translated to an image classification task for providing preliminary status of the tissue before the continuous prediction.

### D. Vision Model and Image Processing

As the knife progresses from 0mm to 35mm, it first travels in free space, then penetrates the tissue model at 12mm. The deformations noticed are classified as 'no', 'minor', 'medium' and 'large'. A larger number of classes can be chosen if the application demands. The camera captures hundreds of images while the knife is at random positions, thus generating images of random deformation levels. The samples and knife positions for the four representing classes are shown in Figure 3. It is observed that deformation is non-linearly related to the knife positions, and hence force estimate alone is insufficient.

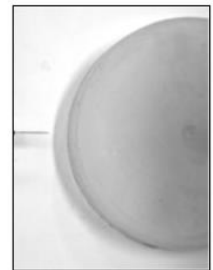

Compress **00**
**Zero** deformation **(0%)**
Knife position : (0-12) mm

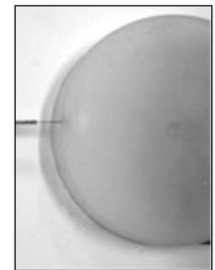

Compress **01**
**Minor** deformation **(33%)**
Knife position : (12-21.7) mm

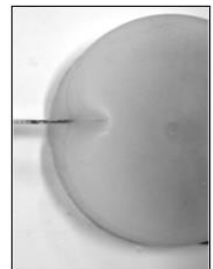

Compress **02**
**Moderate** deformation **(67%)**
Knife position : (21.7-30.1) mm

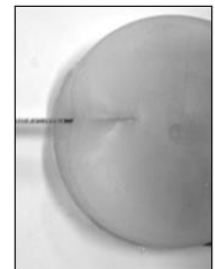

Compress **03**
**Max** deformation **(100%)**
Knife position : (30.1-35) mm

Fig. 3. Knife positions for various deformation levels – nonlinear relationship

With the above reference samples, a vision model is deployed to train 1500 random images. Each image is pre-classified to a deformation level and pre-processed to generate an even larger dataset for training. The image processing technique crops the original image to retain only the relevant information. The full resolution images then go through image argumentation with ratio/brightness change, flips, rotations, color channel shifts, etc to create more possible examples for the training data set (Figure 4).

### E. Classification Model

To train the dataset and classify the images, YOLO model, which frames object detection as a single regression problem (straight from image pixels to class probabilities), is employed. This approach diverges from traditional methods that repurpose classifiers or localizers to perform detection and apply the model to an image at multiple locations and scales. YOLO's key advantage is its speed, thus processing images in real-time, achieving high frame rates without sacrificing accuracy. The success of YOLO model in vision-based classification is shown in [14]. YOLOv8m has been utilized in this work, owing to its better backbone networks, anchor-free detection heads, and improved training strategies [15]. These enhancements make YOLOv8 faster and accurate, making it particularly suitable for surgical robotic application, where timely and precise detection is crucial. Some key hyperparameters for the model training are listed in 'SupplemantaryMaterial'. The dataset available after image processing is divided into training-evaluation-testing set in a 7:2:1 ratio.

### F. Regression Model:

The object detection model could nonetheless force classify an image even if it does not belong to any particular class. To overcome this limitation, the regression model is used to output a continuous prediction of the deformation. The exact deformation is essential for the surgeon's clinical assessment, in addition to the classification of compression level. A CNN regression model is trained for this purpose. A notable work which utilizes vision-based solution with supervised learning can be seen in [16], and CNN based approaches for estimating the deformation are already presented in [17]. However, a novel approach is presented here where contour-based numerical data extracted from images is utilized as target data. This is more realistic for the surgical robotic scenario.

The model detects edges in the image, and highlights areas where the intensity of pixel values changes rapidly, which potentially correspond to the object boundaries or deformations (detailed with figures in Section III-C). The area enclosed by each detected contour is computed, with contour areas corresponding to degrees of deformation. The deformed image set is collected, and a numerical target value is assigned to each image based on the deformation level. A fully connected CNN regression model that outputs numerical data is then employed.

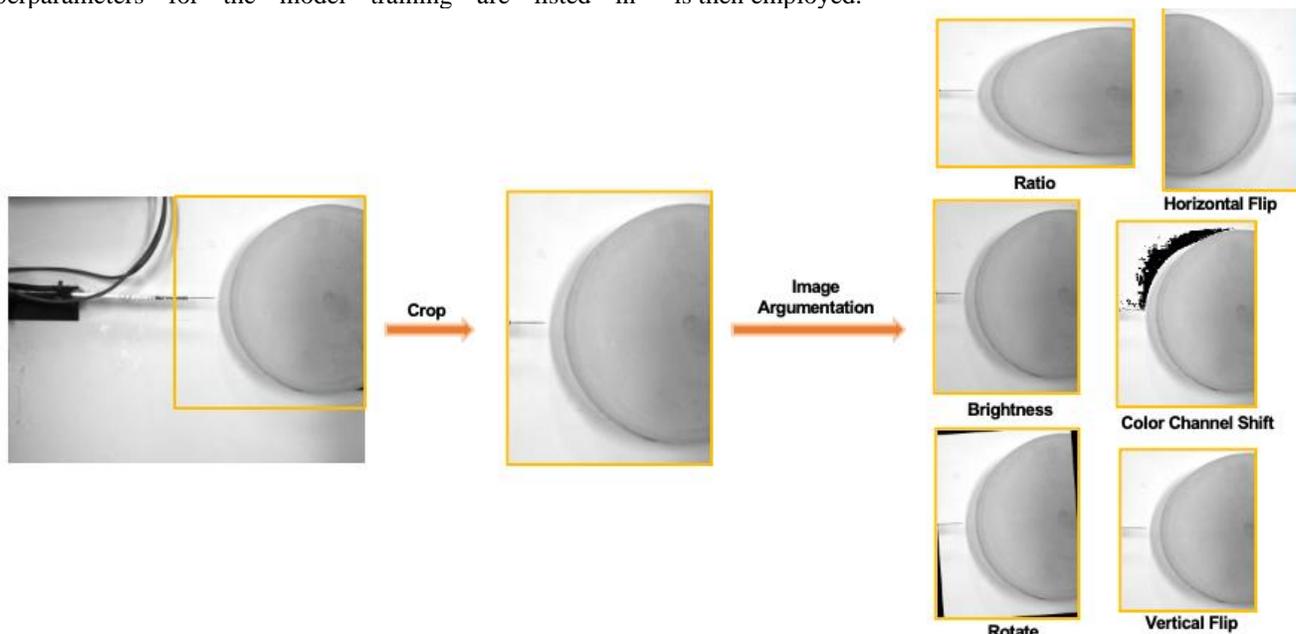

Fig. 4. Image processing upon a sample image – expanding the dataset by creating variants of the original saves time complexity

## III. RESULTS

### A. Force Estimation from Current Data

With the experimental setup described in Section II-A, studies conducted with two tissue samples are discussed.

#### 1). Softer tissue model

A sample of tissue model is made with EcoFlex10 and the surgical knife is allowed to penetrate it. The total range of for the knife to completely probe is 35mm (20,000 encoder increments). At an instant of t=3.42s, the knife is commanded to probe, and reaches its target at t=3.77s. The knife is let to stay inside the tissue model for a brief period (to notice any possible changes in the force) and is commanded the backward journey at t=6.33s, reaching its original position at t=6.72s.

The results obtained during this interaction are shown in Figure 5. These experiments are conducted by letting the motor shaft rotate at a maximum speed of 200rpm, while having quite high acceleration and deceleration of 20000rpm/s.

### 2). Harder tissue model

Another test is conducted with EcoFlex30, which is of higher stiffness. Characteristics of both the materials used are reported in 'SupplementayMaterial'. The experimental parameters all remain the same as described in Section III-A, while only the time of command differing. The position and force profiles during this interaction are reported in Figure 5, and discussed in Section III-A.

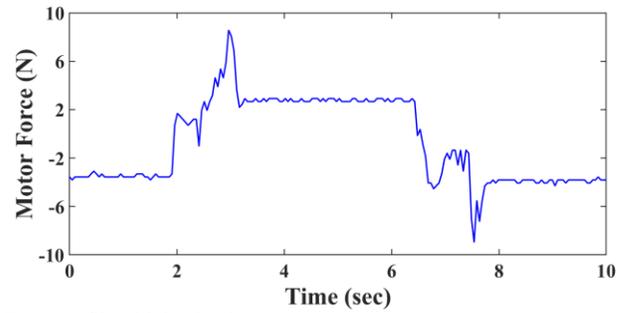

d) Force profile with harder tissue

Figure 5. Position and force profiles during tool-tissue interaction

## B. External Force Sensing

The measurement of the forces read by the sensor for the experiment conducted with EcoFlex 10 (softer tissue) are reported in Figure 6. For comparison, the forces estimated from the motor current are also overlapped. The forces read by the sensor in other two perpendicular directions and the moments in all three directions are also reported.

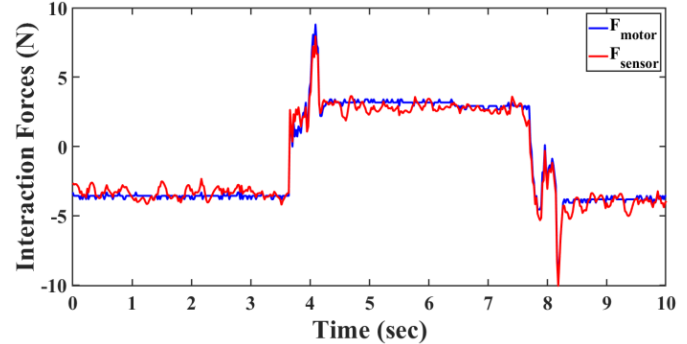

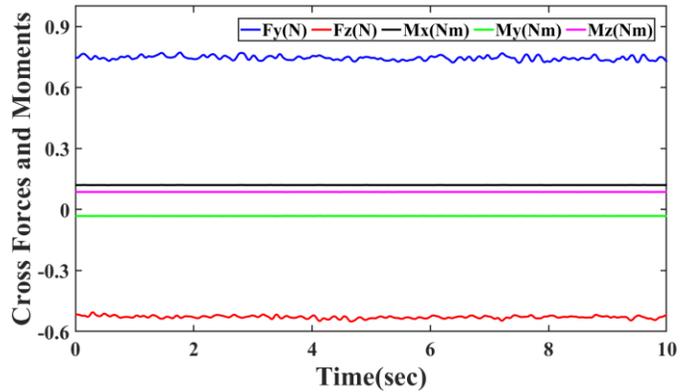

Fig. 6. External force sensor measurements with softer tissue

A low pass filter with a cut-off frequency of 0.1Hz is employed on the output of the force sensor to eliminate the noises, making the smoothened signal suitable for comparison. The forces obtained through this external sensing method are very close to the estimates from raw data. For enhanced accuracy, these two measurements are fused using a Kalman filter, to eliminate any possible noises. It is also an established fact that the output from Kalman filter produces a covariance lesser than *each* of the original input signals, meaning the actual possibility of deviation error is minimized compared to both the individual measurements [18].

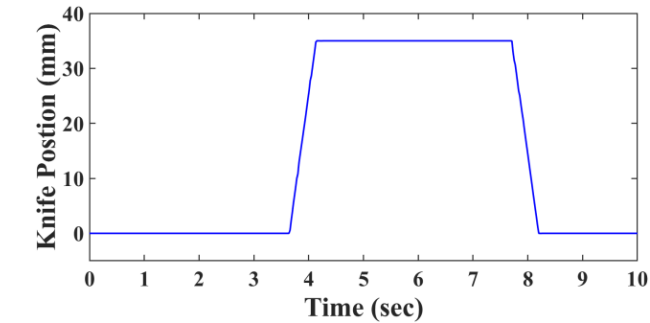

a) Position profile with softer tissue

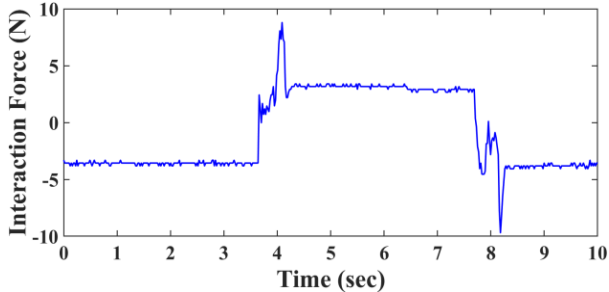

b) Force profile with softer tissue

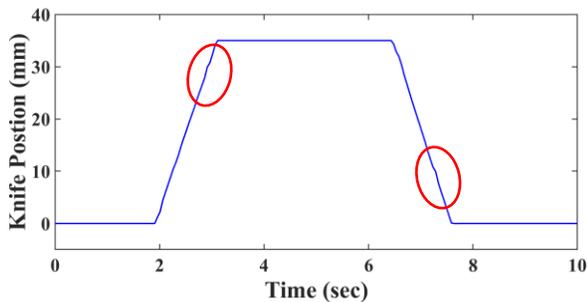

c) Position profile with harder tissue

## C. Tissue Classification and Deformation Estimation

100 images of each compression level are tested using the YOLO and regression models. For a sample image in Compress01 category, the contour development with the regression model is shown in Figure 7(a). The counter is used to first calculate counter areas and then project to deformation levels. Secondly, the YOLO prediction probabilities for each class are obtained as 0.0213, 0.9783, 0.0004 and 0 respectively. From these raw deformation probabilities, an optimized deformation is computed, which would be particularly useful when the probabilities of two higher classes are comparable, and deformation does not fall into an explicit trained class. Details on the optimization methodology and the deformation estimate for untrained classes, (deformation levels outside defined classes' range) are discussed in 'SupplementaryMaterial'.

$$Optimized\_Deformation=(0.9783*33\%)+(0.0213*0\%)+(.004*67\%)+(0*100\%)=32.5519\%$$

In this case study, the deformation estimate from the YOLO model is almost the Compress01's measure of 33%. The obtained raw and optimized deformation estimates can be visualized in Figure 7(b).

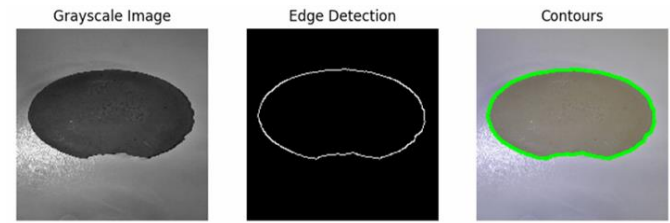

(a) Contour development for computing deformation area

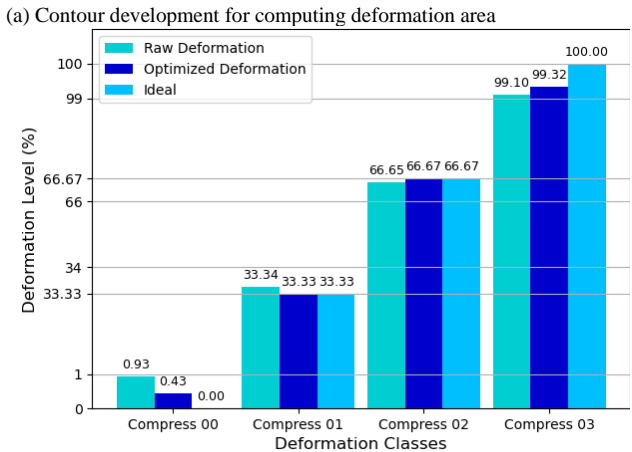

*Raw: Estimate from HIGHEST probability prediction*
*Optimized: Formulated from combining ALL top probabilities*
*Ideal: Training dataset*

(b) Deformation prediction for each compression level and Accuracy comparison

Fig. 7. Case study of a sample image in Compress01

The area enclosed by the contours is calculated and the data set for the CNN regression model is split into 7:2:1 for training, validation, and testing. The predicted deformation areas for the test set are reported in Figure 8.

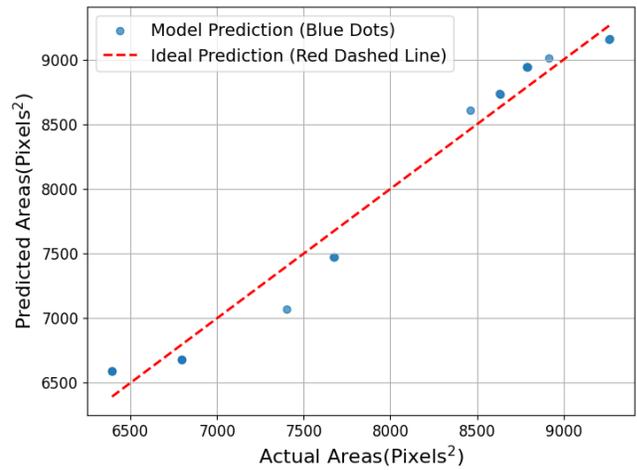

*A lower contour area indicates higher deformation.*
Fig. 8. Predicted deformation area – comparison with numerical estimate

The predictions are observed to be quite accurate, with an RMSE of 4.23%.

## IV. DISCUSSION

### A. Force Estimation

It can be observed from Figure 5 that the position plot is very smooth. This demonstrates the precision of the controller, maintaining the supplied position profile irrespective of any obstacles such as the tissue. Indeed, penetration into the tissue model occurred at a position of 13mm, but there are no interruptions seen in the position profile at this position.

The force profile of softer tissue shows interesting results, which is the prime interest in this study. The force drawn by the motor spikes when it starts from rest, and the interaction force can be noted during the period *t=3.7-6.6s*. It is inferred that an average of *2.26N* was demanded, which is realistic to the human tissue [10].

Interesting facts can be noted from the position profile with harder tissue. Unlike the previous case, the position plot has very minor interruptions (circled in red) in its smooth flow at instances *t=2.9s* (while entering) and *t=7.28s* (while leaving the tissue). This is expected as the tissue model now offers a higher resistance compared to the previous case of EcoFlex 10. Corroborating to this fact, the time taken for probing is also noted as *1.21s*, which is 145% greater than the earlier case of *0.35s*.

In the force profile of harder tissue, the change in current drawn from the rest phase of the knife to the probing phase is noted as 6.51N, which is 39% higher than the 4.66N observed with EcoFlex10.

### B. Deformation Estimation

The raw deformation predication yields a close correlation with the ideal result, while prediction applying optimization shows an even closer match. The regression model's output can be cross-verified with the class identified by YOLO. This mode of classification of tissue compression level is proven [19] to be highly effective in identification of the disease condition. Such analysis when carried on pretumoral tissues also provides early evidence of tumor recurrence.

V. Conclusion

A means through which surgeon's perception levels can be enhanced in robotic surgery is presented in this paper. Force estimation is performed from the raw data to minimize errors and is augmented with external sensing. This provides the surgeon with precise knowledge of the forces applied on the patient. Furthermore, the degree of deformation is computed through which the surgeon can infer the healthiness of the patient. Convolutional neural networks are used to classify the tissue's compression and the presented regression model computes the deformation.

Supplementary Materials

The supplementary material provides more details on the objectives met in this work, positioning of external sensor and camera, LabView interface panel, the hyperparameters used for training CNN, tissue material characteristics, and additional results.

# Supplementary Materials

## Estimation of Tissue Deformation and Interactive Force in Robotic Surgery through Vision-based Learning

Srikar Annamraju, Yuxi Chen, Jooyoung Lim, Inki Kim, *Member, IEEE*

## VI. INTRODUCTION

The Da Vinci robot from Intuitive Surgicals®, Sunnyvale, CA, USA, is pretty much having monopoly in the robotic surgical field. One of the prime concerns still looking for answers during its implementation is the lack of force-feedback in the system.

Surgical robots are currently unable to provide force feedback since such a feedback loop, in the presence of time delay, causes instability in the system. Instability can also affect the forward signal transmission, which is the position command, and eventually causing undamped oscillations in the end-effector of the robot (the surgical tool-tip). Thus, to retain accuracy in position (and consequently patient safety), force feedback has been compromised.

The following are the objectives met in this work:
1. To develop a prototype for robotic tool-tissue interaction
2. To incorporate control for precise motion of the robotic tool, enabling penetrating the tissue
3. To estimate the interactive forces between the tool-tip and the tissue through motor currents
4. To measure the interactive forces through an external sensor and fuse with estimate from currents
5. To classify the tissue based on its deformation level
6. To develop a regressor model for precise continuous estimation of compression level

## VII. MATERIALS AND METHODS

### A. Development of Prototype

Given the 7 degrees of freedom (DoF) complexity of typical surgical robotic systems (such as Da Vinci or RAVEN), it is meaningful to focus clinical assessment of the tissue on a simpler system. Two perpendicular linear motions are chosen on the horizontal plane, such that they span the entire workspace of the operating zone.

### B. External Force Sensing and Fusion

The external force sensor used and its placement on the tool is shown in Figure 1 of this material.

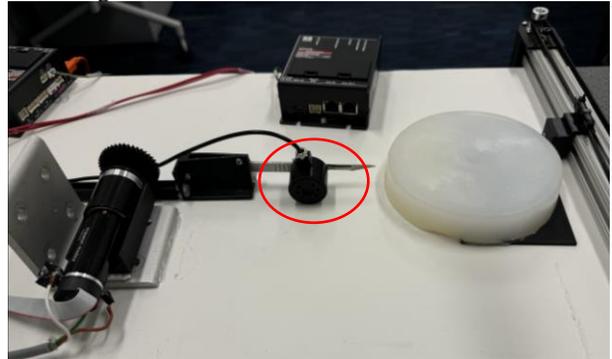

Figure 1. Positioning of the external force sensor (circled in red)

For the interface of the controllers and the sensor, the front panel developed in LabView® is shown in Figure 2 of this material.

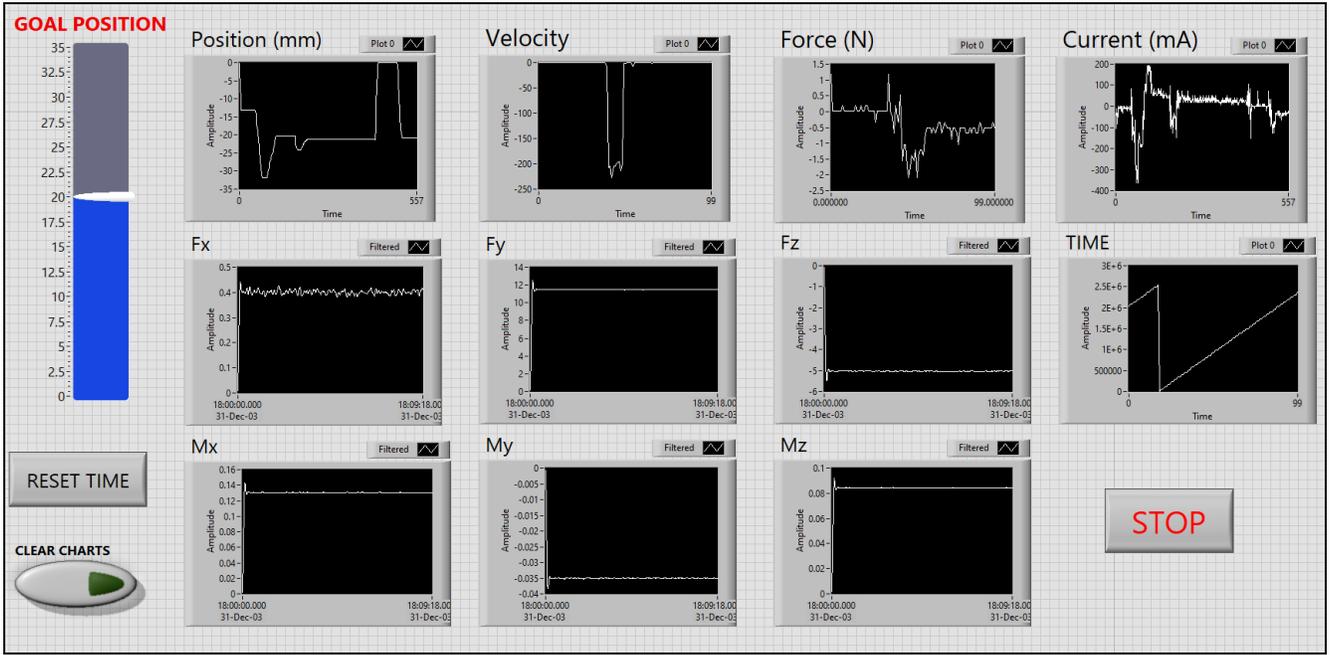

Fig. 2. LabView front panel of implementation

### C. Camera Integration

The camera described in the main body manuscript captures the top-view of the tool-tissue interaction, and it's placement can be seen in Figure 3 of this material.

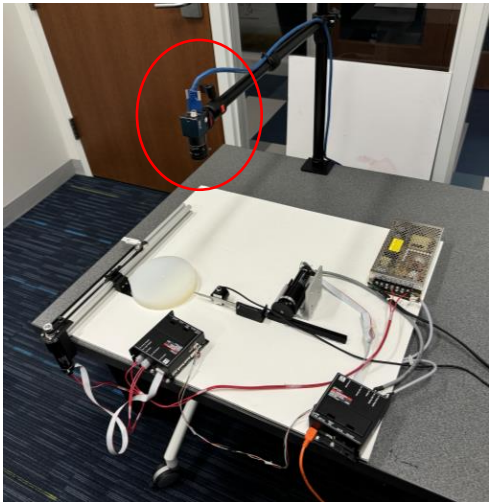

Fig. 3. Camera Positioning (circled in red)

### D. Classification Model

The YOLOv8 model is detailed in the paper as a tool for classification of tissues. The hyperparameters used for training the model are shown in Figure 4 of this material.

```
img_size: 640
batch_size: 16
epochs: 100
learning_rate: 0.01
weight_decay: 0.0005
momentum: 0.9
warmup: 950
optimizer: 'SGD'
lr_scheduler: 'cosine'
flip_lr: 0.5
color_jitter: 0.25
random_crop: 0.3
conf_threshold: 0.25
nms_threshold: 0.45
iou_threshold: 0.5
```

Fig. 4. Hyperparameters used for YOLOv8m training

## VIII. RESULTS

### A. Force Estimation from Current Data

As explained in the main document, results are reported when the robotic tool interacts with two distinct tissue models. A comparison of the dynamic properties of these two materials is shown in Table I of this material.

TABLE 1. Comparison of EcoFlex tissue model properties

|  | EcoFlex 10 | EcoFlex 30 |
|---|---|---|
| **Viscosity** | 14,000cps | 3,000cps |
| **Tensile strength** | 120psi | 200psi |
| **Elongation** | 800% | 900% |
| **Modulus** | 8psi | 10psi |

*B. Tissue Classification and Deformation Estimation*

While both YOLO and regression models are used in this paper, a comparison between these models, is shown in Table II of this material. A novel multimodal approach is suggested as a future work, and its merits are also reported in this table.

TABLE II. YOLO vs. Multimodal Model vs. Image Regression Model

| MODEL TYPE | ADVANTAGE | DISADVANTAGE |
|---|---|---|
| YOLOv8 | • Fast and efficient: Optimized for real-time performance.<br>• Good for classification tasks: Works well with predefined categories.<br>• Easy to fine-tune for custom datasets. | • Limited to discrete classes: Unable to model continuous deformation.<br>• Forced classification: Assigns one category even if the image does not belong to any categories.<br>• No regression capability: Cannot estimate the degree of deformation as a continuous value. |
| CNN Regression Model with Contour numerical data | • Simpler architecture: Uses only images to estimate continuous deformation values.<br>• Continuous predictions: Provides smooth, quantitative estimates (e.g., deformation percentage).<br>• Quantitative insights via contour-based features such as deformation area and edge density | • May overfit to specific visual patterns if the dataset is not diverse enough.<br>• Accuracy may be limited: Relies solely on visual features, which might not capture all influencing factors.<br>• Computational cost may increase with large datasets and contour calculations. |
| Multimodal Model | • Integrates multiple data sources: Combines image data with physical data (e.g., force feedback).<br>• Better accuracy: Leverages richer information for more reliable predictions.<br>• Handles both regression and classification tasks: Suitable for continuous deformation estimation. | • Requires additional data: Needs data from sensors or external devices.<br>• Higher complexity: More challenging to implement and train compared to single-input models.<br>• Slower inference speed: Due to the combination of multiple data streams. |

A brief guideline for choosing the model for a given application is presented herein:

*(i) YOLO (Image Only):* Useful when speed is critical and the task only requires classification into discrete categories (e.g., Compress00–Compress03). Suitable for real-time monitoring but limited for continuous deformation estimation.

*(ii) Image Regression Model:* If a simple, continuous prediction of deformation (e.g., percentage of deformation) using only image data is needed. Ideal when external sensor data is unavailable, but a quantitative estimate is still required.

*(iii) Multimodal Model:* If access to additional sensor data (e.g., scalpel force, position) is available and needed to enhance prediction accuracy by combining multiple data streams. Best for detailed analysis and scenarios where deformation estimation depends on multiple factors.

*1) Optimization Method*

The optimization method followed in the paper is explained here:

If the probability of the top1 class is > 90% for *Compress00* OR > 95% for *Compress01/02/03*, the deformation level is precisely in the trained class, and the corresponding standard deformation (0%/33%/67%/100%) can be taken as the prediction.

If not, the probability of the top2 classes are taken and the optimized deformation is applied to calculate the deformation prediction of the image.

*2) Deformation on Untrained Class*

A series of images of the knife gradually penetrating the tissue are recorded. For images with deformation levels not in the trained categories, some patterns are observed. When the images show deformation leaning heavily toward a trained category, the model still gives a high probability for that category. However, changes in deformation percentage Prediction are observed, albeit non-linearly. For example, in Figure 5 of this material, the yellow-boxed image shows a significant change in deformation level, but the deformation percentage only rises by 1%. When the deformation level falls between two trained categories, the model appropriately predicts the deformation level. For example, in the image below, the last image accurately shows an appropriate deformation level (~47%).

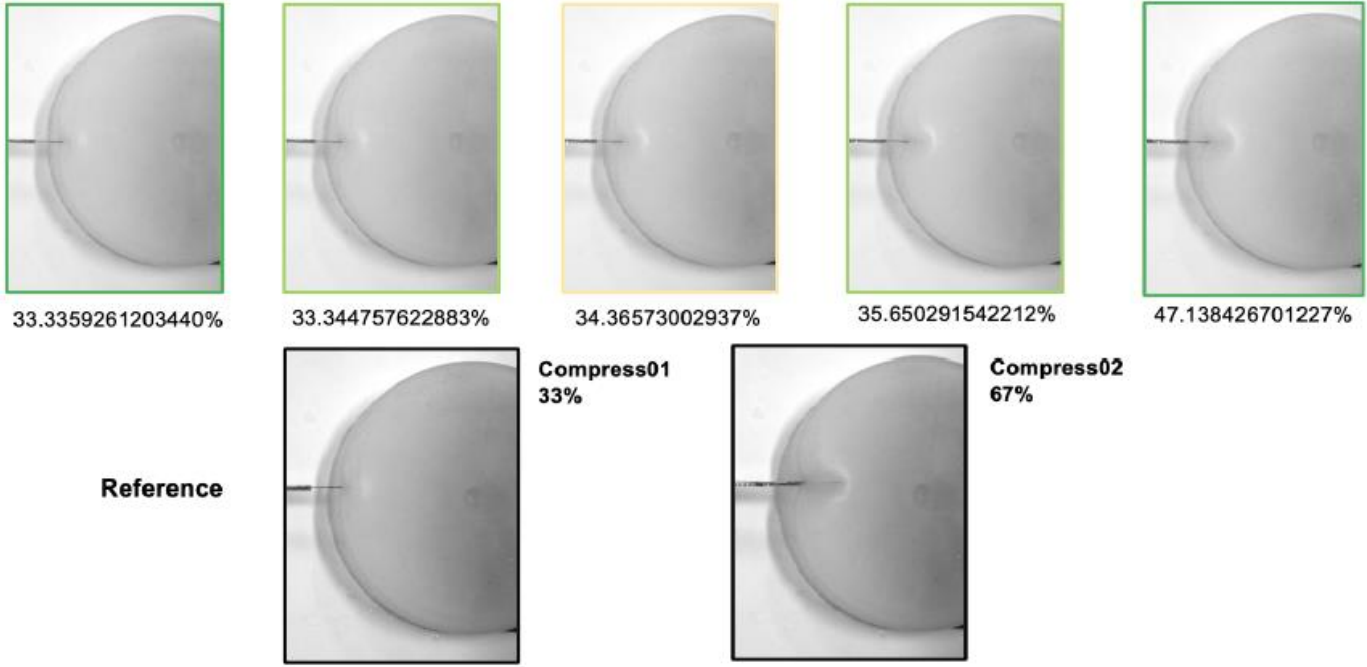

Fig. 5. Deformation Prediction on Untrained Images

The model can classify images into more deformation categories beyond the four initially trained. However, using linear probability to calculate deformation is imprecise. The relationship between calculated deformation (based on probability) and actual deformation needs to be determined, or a nonlinear method to transform probabilities into deformation levels needs to be developed. Focusing more on the second-highest probability when determining deformation can be considered.

## UNITS AND DIMENSIONS

N    – Newtons (Units of Force)
Nm   – NewtonMeter (Units of Torque/Moment)
cpt  – Counts Per Turn (Units for encoder increments)
s    – Seconds (Units for Time)
MP   – Mega Pixel (Units for Camera resolution)
V    – Volts (Units for Voltage)
A    – Amperes (Units of Current)
Hz   – Hertz (Units of Frequency)
CAN – Controller Area Network (A communication protocol)
YOLO – You Only Look Once (Object detection algorithm)